\documentclass[twocolumn,showpacs,aps,prl,superscriptaddress]{revtex4}

\usepackage{graphicx}
\usepackage{dcolumn}
\usepackage{amsmath}
\usepackage{epsfig}

\input{pubboard/babarsym}

\newcommand{\BABARPubNumber}  {09/013}

\newcommand{\SLACPubNumber} {13673}
\newcommand{\LANLNumber} {0906.2219}

\newcommand{\eg}{$E_{\gamma}$}
\newcommand{\threeS}{$\Upsilon(3S)$}
\newcommand{\Apl}{${\cal{A}}_{pl}$}
\newcommand{\ttoe}{$\tau^+\rightarrow e^+\nu_e \overline{\nu}_{\tau}$}
\newcommand{\ttom}{$\tau^+\rightarrow \mu^+ \nu_{\mu} \overline{\nu}_{\tau}$}

\def\figurebox#1#2#3{
    \def\arg{#3}
    \ifx\arg\empty
    {\hfill\vbox{\hsize#2\hrule\hbox to #2{\vrule\hfill\vbox to #1{\hsize#2\vfill}\vrule}\hrule}\hfill}
    \else
    {\hfill\epsfbox{#3}\hfill}
    \fi}

\long\def\inst#1{\par\nobreak\kern 4pt\nobreak
    {\it #1}\par\vskip 10pt plus 3pt minus 3pt}

\begin{document}

\preprint{\babar-PUB-\BABARPubNumber} 
\preprint{SLAC-PUB-\SLACPubNumber} 

\begin{flushleft}
\babar-PUB-\BABARPubNumber\\
SLAC-PUB-\SLACPubNumber\\
hep-ex/\LANLNumber\\ [10mm]

\end{flushleft}
\title{ {\large {\bf \boldmath Search for a low-mass Higgs boson in
$\Upsilon(3S)\rightarrow\gamma A^0$, $A^0\rightarrow\tau^+\tau^-$ at
\babar\/} } }

%
\author{B.~Aubert}
\author{Y.~Karyotakis}
\author{J.~P.~Lees}
\author{V.~Poireau}
\author{E.~Prencipe}
\author{X.~Prudent}
\author{V.~Tisserand}
\affiliation{Laboratoire d'Annecy-le-Vieux de Physique des Particules (LAPP), Universit\'e de Savoie, CNRS/IN2P3,  F-74941 Annecy-Le-Vieux, France}
\author{J.~Garra~Tico}
\author{E.~Grauges}
\affiliation{Universitat de Barcelona, Facultat de Fisica, Departament ECM, E-08028 Barcelona, Spain }
\author{M.~Martinelli$^{ab}$}
\author{A.~Palano$^{ab}$ }
\author{M.~Pappagallo$^{ab}$ }
\affiliation{INFN Sezione di Bari$^{a}$; Dipartimento di Fisica, Universit\`a di Bari$^{b}$, I-70126 Bari, Italy }
% ...
\collaboration{The \babar\ Collaboration}
\noaffiliation

\date{\today}
\begin{abstract}
We search for a light Higgs boson, $A^0$, in the radiative decay
$\Upsilon(3S)\rightarrow\gamma A^0$, $A^0\rightarrow\tau^+\tau^-$,
\ttoe\ or \ttom\/.  The data sample contains 122 million
$\Upsilon(3S)$ events recorded with the \babar\ detector. We find no
evidence for a narrow structure in the studied $\tau^+\tau^-$
invariant mass region of $4.03<m_{\tau^+\tau^-}<10.10$ \gevcc\/. We
exclude at the $90\%$ confidence level (C.L.) a low mass Higgs
decaying to $\tau^+\tau^-$ with a product branching fraction ${\cal
{B}}(\Upsilon(3S)\rightarrow\gamma A^0)\times {\cal
{B}}(A^0\rightarrow\tau^+\tau^-)$ $>(1.5-16)\times 10^{-5}$ across the
$m_{\tau^+\tau^-}$ range. We also set a 90$\%$ C.L. upper limit on the
$\tau^+\tau^-$-decay of the $\eta_b$ at
${\mathcal{B}}(\eta_b\rightarrow \tau^+\tau^-)<8\%$.
\end{abstract}
\maketitle

In the standard model (SM) of particle physics~\cite{ref:stdmodel},
fundamental particles acquire mass via the Higgs
mechanism~\cite{ref:higgs}. This mechanism requires the existence of
at least one new particle called the Higgs boson. In the SM, there is
only a single Higgs boson, with a mass of the order of the electroweak
unification scale ($\sim 100$ \gevcc\/). In the minimal supersymmetric
standard model (MSSM), additional Higgs doublets are required to give
mass to the new particles~\cite{Haber:1984rc}. Moreover, in the
next-to-minimal supersymmetric standard model (NMSSM), an additional
Higgs singlet field is introduced to solve the hierarchy
problem~\cite{ref:dgm}. A linear combination of this singlet with a
Higgs doublet leads to a {\it CP}-odd Higgs state, $A^0$, whose mass
need not be larger than $2m_b$, where $m_b$ is the $b$-quark
mass~\cite{ref:dgm,Hiller:2004ii}. It is ideal to search for this
state in $\Upsilon\rightarrow\gamma A^0$ decays~\cite{ref:wilcek}. The
branching fraction ${\mathcal{B}} (\Upsilon(3S)\rightarrow \gamma
A^0)$ depends on the NMSSM parameters, but a value as large as
$10^{-4}$ is plausible for reasonable parameters~\cite{ref:dgm}. In
the mass range where the decay $A^0\rightarrow\tau^+\tau^-$ is
kinematically accessible, this mode is expected to
dominate. Constraints on the invisible~\cite{BAD2073} and
dimuon~\cite{Aubert:2009cp} decays of the $A^0$ have recently been
obtained.

The current best limit on the product of branching fractions
${\cal{B}}(\Upsilon(1S)\rightarrow\gamma A^0)\times {\cal
{B}}(A^0\rightarrow\tau^+\tau^-)$ is given by the CLEO
Collaboration~\cite{ref:cleo} based on a data sample of 21.5 million
$\Upsilon(1S)$ candidates. The CLEO $90\%$ C.L. limits cover the range
$2m_{\tau}<m_{A^0}<9.5$ \gevcc\ ($m_{\tau}$ is the $\tau$-lepton
mass~\cite{Yao:2006px}) and vary between $1\times 10^{-5}$ and
$48\times 10^{-5}$. A recent D0 search for a neutral pseudoscalar Higgs 
boson in a similar mass range showed no significant 
signal~\cite{Abazov:2009yi}.

In this Letter, we study the decays
$\Upsilon(3S)\rightarrow\gamma\tau^+\tau^-$, where the search for
$A^0$ is extended for a wider mass range w.r.t the
$\Upsilon(1S)\rightarrow\gamma \tau^+\tau^-$.  We scan for peaks in the
distribution of the photon energy, $E_\gamma$, corresponding to peaks
in the $\tau\tau$ invariant mass $m^2_{\tau^+\tau^-} = m^2_{3S}
-2m_{3S}E_\gamma$, where $m_{3S}$ is the \threeS\ mass and \eg\ is
measured in the \threeS\ rest frame (center-of-mass (CM) frame). We
quote branching fraction values in the region $4.03 < m_{\tau^+\tau^-}
< 10.10$ \gevcc\/, but we exclude from our search the region $9.52 <
m_{\tau^+\tau^-} < 9.61$ \gevcc\/, because of the irreducible
background of photons produced in the decay chain
$\Upsilon(3S)\rightarrow\gamma\chi_{bJ}(2P)$,
$\chi_{bJ}(2P)\rightarrow\gamma\Upsilon(1S)$, where $J=0,1,2$. In
addition, we set an upper limit on
$\mathcal{B}(\eta_b\rightarrow\tau^+\tau^-)$.

The data were collected with the \babar\ detector~\cite{ref:detector}
at the PEP-II asymmetric-energy $e^+e^-$ storage rings at the SLAC
National Accelerator Laboratory, operating at the $\Upsilon(3S)$
resonance. We use a data sample of 122 million \threeS\ events,
corresponding to an integrated luminosity of 28 fb$^{-1}$. We also use
data samples of $2.6$ fb$^{-1}$ recorded 30 \mev\ below the \threeS\
(OFF3S), 79 fb$^{-1}$ at the $\Upsilon(4S)$ (ON4S), and 8 fb$^{-1}$ 40
\mev\ below the \FourS\ resonance (OFF4S) to study the background and
to optimize the selection criteria. These data samples were taken with
the same detector configurations. Monte Carlo (MC) event samples based
on {\textsc{Geant4}}~\cite{ref:geant} simulation of the detector are used
to optimize selection criteria and evaluate efficiencies.

We select events in which both $\tau$-leptons decay leptonically,
\ttoe\ or \ttom\ (denoted in the following as $\tau\rightarrow e$, or
$\tau\rightarrow \mu$)~\cite{charge}. Events are required to contain
at least one photon with \eg\/$>100$ \mev\/, and exactly two charged
tracks. We allow up to nine additional photons with energies below 100
\mev\/ in the CM frame. Photons are reconstructed from localized
deposits of energy in the electromagnetic calorimeter, which have
energies larger than 50 \mev\ in the laboratory frame and which are
not associated with a charged track. Both charged tracks are required
to be identified as leptons ($e$ or $\mu$). After this selection the
residual background is mostly due to
$e^+e^-\rightarrow\gamma\tau^+\tau^-$ and higher order QED processes,
including two-photon reactions such as $e^+e^-\rightarrow
e^+e^-e^+e^-$ and $e^+e^-\rightarrow e^+e^-\mu^+\mu^-$ with smaller
contributions from other $\Upsilon(3S)$ decays and $e^+e^-\rightarrow
q\bar{q}$ ($q=u,d,s,c$).

To reduce this residual background, we exploit a set of eight
discriminating variables: the total CM energy ($E_{\mathrm{total}}$)
calculated from the two leptons and the most energetic photon; the
squared missing mass ($m^2_{\mathrm{miss}}$) obtained from the missing
four-momentum, which is the difference between the final and initial
state momenta; the aplanarity (\Apl\/), which is the cosine of the
angle between the photon and the plane of the leptons; the largest
cosine between the photon and one of the tracks
($\cos\theta_{\gamma\mathrm{-track}}$); the cosine of the polar angle
of the highest-momentum track ($\cos\theta_{\mathrm{track}}$); the
transverse momentum of the event ($p_T$) calculated in the CM frame;
the cosine of the polar angle of the missing momentum vector
($\cos\theta_{\mathrm{miss}}$); and the cosine of the opening angle
between the tracks in the photon recoil frame
($\cos\theta_{\mathrm{open}}$). The final selection criteria on these
variables are obtained by maximizing the quantity $S/\sqrt{B}$, where
$S$ ($B$) stands for the expected number of signal (background)
events. Numbers of signal events are obtained from MC samples, while
background yields are obtained from the OFF3S, ON4S, and OFF4S
datasets. Since the background varies as a function of the photon
energy, we optimize the selection criteria in five \eg\ regions:
($S_1$) $0.2<E_{\gamma}<0.5$ \gev\/, ($S_2$) $0.5<E_{\gamma}<2.0$
\gev\/, ($S_3$) $1.5<E_{\gamma}<2.5$ \gev\/, ($S_4$)
$2.5<E_{\gamma}<3.5$ \gev\/, and ($S_5$) $3.0<E_{\gamma}<5.0$
\gev\/. The overlaps between the \eg\ regions reduce the discontinuity
in the efficiency at the boundaries. The dominant irreducible
background is due to $e^+e^-\rightarrow \gamma\tau^+\tau^-$. The
highest level of background contaminations is observed at low \eg\
values. Among the different final states, the background is largest in
$\tau\tau\rightarrow ee$ and smallest in $\tau\tau\rightarrow e\mu$.

The photon energy resolution degrades as a function of \eg\/, from 8
\mev\/ at \eg\/$\sim 0.2$ \gev\ to 55 \mev\ at \eg\/$\sim 4.5$
\gev\/. The selection efficiency is calculated using MC events. The
efficiency in the $\tau\tau\rightarrow ee$, $\tau\tau\rightarrow
e\mu$, and $\tau\tau\rightarrow \mu\mu$ modes varies as a function of
\eg\/ between 10--14$\%$, 22--26$\%$, and 12--20$\%$,
respectively. The MC samples are generated with angular decay
distributions expected for a {\it CP}-odd Higgs boson; similar
efficiencies are obtained for {\it CP}-even states.

We search for an excess in a narrow region in the \eg\ spectrum since
any peak in the recoil mass ($m_{\tau\tau}$), indicating the presence
of a new particle decaying in $\tau$-pairs, translates to a peak in
the \eg\ distribution. We describe the \eg\ distribution as a smooth
background spectrum and a narrow enhancement of known width, but
unknown position and event yield. We perform a binned maximum
likelihood fit simultaneously to the $\tau\tau\rightarrow ee$, $e\mu$,
and $\mu\mu$ samples.

The fit is performed in two steps.  First, we assume there is no
signal and fit the background function. Theoretical
motivations~\cite{Voloshin:2002mv} inspired the choice of the
background function shape, $f=\left
(p(1-x)^r/E^q_\gamma+s/E^5_{\gamma}\right )\cdot
\beta(x)\cdot(3-\beta^2(x))$, where
$\beta(x)\equiv\sqrt{1-4m^2_\tau/(m^2_{3S}(1-x))}$, $x\equiv 2E_\gamma
/m_{3S}$. For each $\tau\tau$-decay mode, a different set of the
parameters $p,q,r,s$ is used. These parameters are allowed to vary.

\begin{figure}[!htbp]
  \begin{center}
    \includegraphics[width=8.5cm]{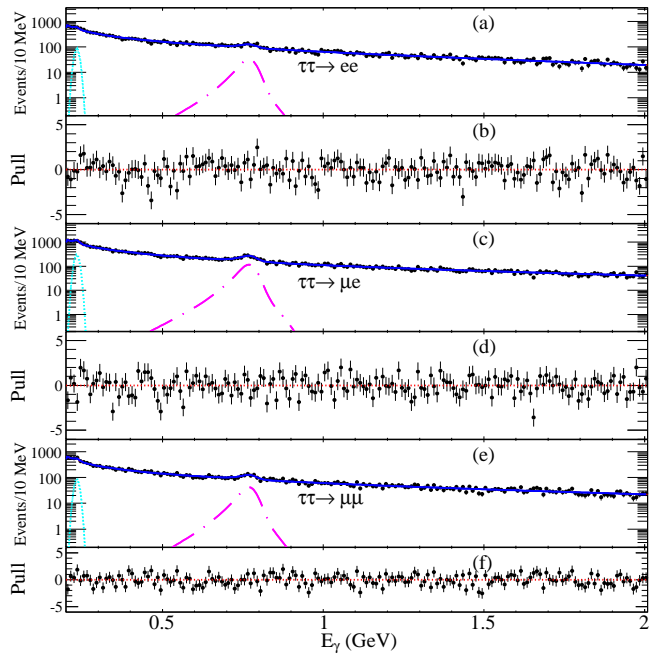}
    \caption{(a), (c), (e): \eg\ distributions for the different
$\tau\tau$-decay modes. Filled circles show the data; dotted
lines represent contributions from
$\Upsilon(3S)\rightarrow\gamma\chi_{bJ}(2P)$,
$\chi_{bJ}(2P)\rightarrow\gamma\Upsilon(2S)$; dotted-dashed lines
show contributions from
$\Upsilon(3S)\rightarrow\gamma\chi_{bJ}(2P)$,
$\chi_{bJ}(2P)\rightarrow\gamma\Upsilon(1S)$; and solid lines show
the total background function. For each $\tau\tau$-decay mode, the
difference between the background function and the data divided by the
uncertainty in the data is shown ((b), (d), (f)).}
    \label{fig:FitForScan}
\end{center} 
\end{figure}

The events $\Upsilon(3S)\rightarrow\gamma\chi_{bJ}(2P)$,
$\chi_{bJ}(2P)\rightarrow\gamma\Upsilon(nS)$, and
$\Upsilon(nS)\rightarrow \tau^+\tau^-$ ($J=0,1,2$; $n=1,2$) are
expected to peak in \eg\ when the photon from
$\chi_{bJ}(2P)\rightarrow\gamma\Upsilon(nS)$ is misidentified as the
radiative photon from the $\Upsilon(3S)$ decay. Each of the peaks in
the photon spectrum due to the
$\chi_{bJ}(2P)\rightarrow\gamma\Upsilon(1S)$ transitions is described
by a Crystal Ball~\cite{ref:cbf} (CB) function. The mean values for
the $\chi_{b0}(2P)$ and $\chi_{b1}(2P)$ CB functions are fixed to the
PDG~\cite{Yao:2006px} and the width values are fixed to the MC
resolution, while the mean and width for $\chi_{b2}(2P)$ are free. The
power law and the transition point for all CB functions used in the
analysis are fixed to the values obtained in MC. The event yields for
the $\chi_{bJ}(2P)$ background for each of the three $\tau\tau$ data
samples are related via their relative efficiencies, which are
functions of \eg\/. To account for the contributions from
$\chi_{bJ}(2P)\rightarrow \gamma\Upsilon(2S)$, a fourth CB function is
added, for which the mean, width, and the relative normalization are
free.  The fitted mean and width obtained for this peak are $234\pm 2$
\mev\ and $13.3\pm 2.7$ \mev\/ (statistical uncertainties only),
respectively. The number of events from the $\chi_{bJ}(2P)\rightarrow
\gamma\Upsilon(nS)$ ($n=1,2$) contamination are common between the
different $\tau\tau$-decay modes, and divided between these modes
according to the efficiency sum, $\epsilon^N=\epsilon_{ee} +
2\epsilon_{e\mu} + \epsilon_{\mu\mu}$, where $\epsilon_{ee}$,
$\epsilon_{e\mu}$, and $\epsilon_{\mu\mu}$ are the efficiencies as a
function of \eg\ in the decay modes $\tau\tau \rightarrow ee$, $e\mu$,
and $\mu\mu$, respectively. An example of the fits to the \eg\
distributions in the different $\tau^+\tau^-$-decay modes, obtained
with the selection criteria $S_1$ and fitted in the region
$0.2<E_{\gamma}<2.0$ \gev\/, are shown in
Fig.~\ref{fig:FitForScan}. Satisfactory fits are obtained.

In the second step of the fit procedure, we search for the signal
$\Upsilon(3S)\rightarrow\gamma A^0$, $A^0\rightarrow\tau^+\tau^-$. We
assume the $A^0$ has negligible width~\cite{Fullana:2007uq}, and
parameterize the signal distribution with a CB function. The search
for such a signal is performed by scanning for peaks in the \eg\
distributions in steps that are equal to half the photon-energy
resolution at any chosen value of \eg\/. In total, 307 scan points are
examined. The mean of the signal function is fixed to the photon
energy at the $i^{\mathrm{th}}$ scan point ($E_\gamma^i$). The signal
width is fixed to the value of the photon energy resolution obtained
from the MC simulation. The contribution from each $\tau\tau$-decay
mode to the total number of Higgs candidates is proportional to the
fractional efficiency for a particular mode. The background shape
parameters (including the $\chi_{bJ}$ parameters) are all fixed to the
values determined in the first step of the fit, with the exception of
$p$ and $s$, to allow free background normalization. The number of
free parameters in each fit is seven ($p_{ee}$, $p_{e\mu}$,
$p_{\mu\mu}$, $s_{ee}$, $s_{e\mu}$, $s_{\mu\mu}$, and
$N_{\mathrm{sig}}$), where the subscripts indicate the final state of
the $\tau\tau$-decay modes. When the scan is performed in the regions
$S_3$, $S_4$, and $S_5$, the parameters $s_{ee}$, $s_{e\mu}$, and
$s_{\mu\mu}$ are fixed to zero.

\begin{figure}[!htbp]
  \begin{center}
    \includegraphics[width=5.0cm]{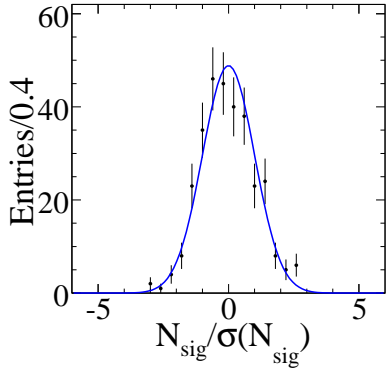}
    \caption{$N_{\mathrm{sig}}/\sigma (N_{\mathrm{sig}})$ as obtained
    from the scanning procedure. Only statistical uncertainties are
    included. The curve shows the standard normal distribution with a
    normalization factor of 307.}
    \label{fig:projection}
\end{center} 
\end{figure}

For each scan point, the yield, $N_{\mathrm{sig}}$, and its
statistical uncertainty, $\sigma(N_{\mathrm{sig}})$ are obtained from
the fit. The yield significance from the data,
$N_{\mathrm{sig}}/\sigma(N_{\mathrm{sig}})$ is shown in
Fig.~\ref{fig:projection}, and overlaid with a standard normal
distribution. The data points are consistent with the normal
distribution, and therefore no significant evidence for any unknown
narrow structure is observed in the scan.

\begin{figure}[!htbp]
  \begin{center}
    \includegraphics[width=8.0cm]{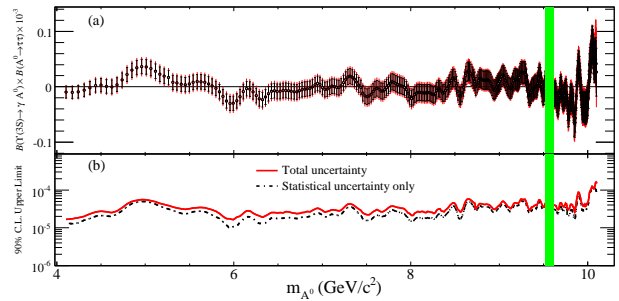}
    \caption{(a) Product branching fractions as a function of the
    Higgs mass. For each point, both the statistical uncertainty (from
    the central value to the horizontal bar) and the total uncertainty
    (statistical and systematic added in quadrature) are shown (from
    the central value to the end of the error bar). In (b), the
    corresponding 90$\%$ C.L. upper limits on the product of the
    branching fractions versus the Higgs mass values are shown, with
    total uncertainty (solid line) and statistical uncertainty only
    (dashed line). The shaded vertical region represents the excluded
    mass range corresponding to the
    $\chi_{bJ}(2P)\rightarrow\gamma\Upsilon(1S)$ states.}
    \label{fig:Newscan_3SAll}
\end{center} 
\end{figure}

Product branching fractions are determined from the signal yields at
each scan point, correcting for a fit bias described below. The
results are shown in Fig.~\ref{fig:Newscan_3SAll}(a). These results
show no evidence for a narrow resonance in the mass range under
study. Bayesian upper limits on the product of branching fractions,
computed with a uniform prior at $90\%$ C.L., are shown in
Fig.~\ref{fig:Newscan_3SAll}(b). The solid line shows the limits
obtained with the total uncertainties (statistical and systematic
added in quadrature) while the dashed line shows the limits with
statistical uncertainties only.

We measure the branching fraction
$\mathcal{B}(\eta_b\rightarrow\tau^+\tau^-)=(-0.1\pm 4.2\pm 2.3) \,
\%$ at $m_{\tau^+\tau^-}=9.389$ \gevcc\/, using the
$\mathcal{B}(\Upsilon(3S)\rightarrow \gamma\eta_b)$ from
Ref.~\cite{:2008vj}. Therefore, the $90\%$ C.L. upper limit on
$\mathcal{B}(\eta_b\rightarrow\tau^+\tau^-)$ is 8 (7)$\%$, considering
all (statistical only) errors and accounting for the expected 10
\mev\/ width of the $\eta_b$. We note that the limit and branching
fraction are insensitive to the $\eta_b$ width within the expected
5-20 \mev\/ range~\cite{:2008vj}.

We account for systematic uncertainties due to tracking ($2\%$),
lepton identification (1.2--2.6$\%$, depending on the $\tau\tau$-decay
mode), photon reconstruction efficiency ($4\%$), and the number of
\threeS\/ ($1\%$). In the scan procedure, the parameters of the
background shape and of the $\chi_{bJ}(2P)$ states are fixed. To
estimate the systematic uncertainty related to these parameters, each
parameter is varied by its estimated statistical uncertainty
determined in the first step of the fit. The scan procedure is
repeated for each parameter change. When calculating the systematic
uncertainties from this source, the correlations between the various
parameters are taken into account. The ratio between the total
systematic uncertainties due to the background shape and the
statistical uncertainties varies between 12$\%$ and 170$\%$. The
largest systematic variations occur for larger values of
$m_{\tau^+\tau^-}$, and are due to the uncertainty in the $q_{e\mu}$
parameter for $\tau\tau\rightarrow e\mu$. The fit bias and its
uncertainty are determined by applying the fit procedure to a large
number of MC experiments. Each MC sample contains a known number of
signal events, while background events are generated according to the
background shape. The event yield, returned by the fit, is a linear
function of the number of input events. The event yield in the data is
corrected using this function. The difference between the corrected
and uncorrected event yield is (conservatively) considered as the
systematic uncertainty due to the fit bias, which is typically small
(few percent) but can be as large as 30$\%$ of the statistical
uncertainty at high $m_{\tau^+\tau^-}$. The systematic uncertainty
associated with the choice of the signal shape function is determined
by varying the values of the parameters in the signal CB function; the
width and the power law are varied (multiplicatively) by $30\%$ and
$38\%$, respectively; the transition point is varied (additively) by
$36\%$. The associated systematic contribution is typically small (few
percent) but is as large as $50\%$ of the statistical uncertainty at
large $m_{\tau^+\tau^-}$.  Finally, we include a systematic
uncertainty of $0.6\%$ to account for the systematic uncertainty due
to the $\tau$ branching fractions~\cite{Yao:2006px}. The dominant
systematic uncertainties are due to the background-shape parameters,
which are obtained from fitting the same data sample. Thus, we
conclude that the main systematic uncertainties are primarily
statistical in nature.

In summary, we have performed a search for a light Higgs boson in the
radiative decays $\Upsilon(3S)\rightarrow\gamma\tau^+\tau^-$, where
\ttoe\ or \ttom\/, using a data sample of 122 million $\Upsilon(3S)$
events. Our search covers the mass range $4.03<m_{\tau^+\tau^-}<10.10$
\gevcc\/, excluding $9.52<m_{\tau^+\tau^-}<9.61$ \gevcc\ to veto the
$\chi_{bJ}(2P)$ with $\chi_{bJ}(2P)\rightarrow \gamma\Upsilon(1S)$. No
evidence for a signature of light Higgs boson decays to $\tau$ pairs
is observed. In this mass interval, the upper limits on the product
branching fraction ${\cal {B}}(\Upsilon(3S)\rightarrow\gamma
A^0)\times {\cal {B}}(A^0\rightarrow\tau^+\tau^-)$ vary between
$(1.5-16)\times 10^{-5}$ at $90\%$ C.L.

We are grateful for the excellent luminosity and machine conditions
provided by our \pep2\ colleagues, 
and for the substantial dedicated effort from
the computing organizations that support \babar.
The collaborating institutions wish to thank 
SLAC for its support and kind hospitality. 
This work is supported by
DOE
and NSF (USA),
NSERC (Canada),
CEA and
CNRS-IN2P3
(France),
BMBF and DFG
(Germany),
INFN (Italy),
FOM (The Netherlands),
NFR (Norway),
MES (Russia),
MEC (Spain), and
STFC (United Kingdom). 
Individuals have received support from the
Marie Curie EIF (European Union) and
the A.~P.~Sloan Foundation.


\begin{thebibliography}{10}
\bibitem{charge} The use of charge conjugate reactions is implied
throughout this Letter.
\end{thebibliography}
\end{document}